\documentclass{ieeeaccess}
\usepackage{cite}
\usepackage{amsmath,amssymb,amsfonts}
\usepackage[capitalise]{cleveref}
\usepackage{algorithmic}
\usepackage{graphicx}
\usepackage{textcomp}
\usepackage{glossaries}
\usepackage{siunitx}
\usepackage{caption}
\usepackage{multirow}
\usepackage{makecell}
\usepackage[caption=false]{subfig}
\usepackage[absolute]{textpos}

\newacronym{dnn}{DNN}{Deep Neural Network}
\newacronym{bnn}{BNN}{Bayesian Neural Network}
\newacronym{pgm}{PGM}{Probabilistic Graybox Model}
\newacronym{sgm}{SGM}{Statistics Graybox Model}
\newacronym{mcdgm}{MCDGM}{Monte-Carlo Dropout Graybox Model}
\newacronym{agf}{AGF}{Average Gate Fidelity}
\newacronym{agif}{AGIF}{Average Gate Infidelity}
\newacronym{mlp}{MLP}{Multi Layer Perceptron}
\newacronym{mle}{MLE}{Maximum Likelihood Estimation}

\newcommand{\control}{\mathbf{\Theta}}
\newcommand{\SX}{\sqrt{X}}

\begin{document}

\history{
This work has been submitted to the IEEE for possible publication. Copyright may be transferred without notice, after which this version may no longer be accessible.
}
\doi{10.1109/TQE.2020.DOI}

\title{Probabilistic Graybox Characterization of Quantum Devices with Bayesian Neural Networks}
\author{
    \uppercase{Poramet Pathumsoot}\authorrefmark{1}, 
    \uppercase{Michal Hajdu\v{s}ek \authorrefmark{1}, and Rodney Van Meter} \authorrefmark{2},
    \IEEEmembership{Senior Member, IEEE}
}
\address[1]{Graduate School of Media and Governance, Keio University Shonan Fujisawa Campus, Kanagawa, Japan (email: poramet@sfc.wide.ad.jp, michal@sfc.wide.ad.jp)}
\address[2]{Faculty of Environment and Information Studies, Keio University Shonan Fujisawa Campus, Kanagawa, Japan (email: rdv@sfc.wide.ad.jp)}
\tfootnote{This work was supported by JST [Moonshot R\&D Program] Grant Number [JPMJMS2061].}

\markboth
{Pathumsoot \headeretal: Preparation of Papers for IEEE Transactions on Quantum Engineering}
{Pathumsoot \headeretal: Preparation of Papers for IEEE Transactions on Quantum Engineering}

\corresp{Corresponding author: Poramet Pathumsoot (email: poramet@sfc.wide.ad.jp).}

\begin{abstract}
    While the Graybox characterization method allows for implicit noise models and is platform-agnostic, the method lacks uncertainty quantification.
    Characterization of quantum devices is a crucial process that enables researchers to gain insight from experimental settings.
    Graybox characterization combines known system dynamics with unknown transformations, where the latter is modeled using machine learning.
    Prediction uncertainty helps researchers make informed decisions. It allows valuable insights from the devices without overconfidence.
    We therefore develop a probabilistic Graybox characterization model using probabilistic machine learning, specifically Bayesian Neural Networks, and utilize binary measurement outcomes directly for inference.
    With stochastic noise in a quantum device, we analyze statistical properties of the measurement data.
    Our results show that the model's prediction performance solely depends on its ability to capture the expected value of the true expectation value.
    Our proposed probabilistic Graybox model outperforms the original model by up to 1.9 times in capturing the distribution of observed data.
    We expect that our results will serve as an additional tool for characterizing quantum devices with uncertainty estimation, as they provide a flexible choice that can be utilized even without extensive prior knowledge of the noise model of the devices.
\end{abstract}


\titlepgskip=-15pt

\maketitle

\section{Introduction} \label{sec:introduction}
Characterization of quantum devices plays a crucial role in the development of quantum technologies. Especially in open-loop optimal control, an accurate \textit{predictive model} of the quantum system is necessary to achieve usable results in experimental settings \cite{machnesTunableFlexibleEfficient2018}.
Even with fault-tolerant quantum error correction (FTQEC), the protocol fails without physical-level control that operates below a noise threshold.
In practice, the error rate might not be constant due to various types of noise, e.g., colored noise, which can potentially affect the trajectory of the quantum state in each execution. When the noise threshold is within the range of physical error uncertainty, FTQEC might fail to improve the logical error rate from the physical error rate \cite{googlequantumaiandcollaboratorsQuantumErrorCorrection2025}.
Constructing a predictive model remains relevant in the presence of closed-loop optimization, as reducing the cost of interacting with the experimental device is also preferred to prevent mishaps due to trial and error, allowing the researcher to utilize the characterized model first.
Researchers can develop a robust, realistic experiment locally without the cost of accessing the quantum device. Thus, accurately characterizing the quantum device is an important step that must be taken carefully.

Constructing a predictive model involves making assumptions about the physical system being modeled. Many approaches assume a closed-form of a Hamiltonian parametrized with system parameters \cite{dasilvaPracticalCharacterizationQuantum2011,huangPredictingManyProperties2020, akhtarScalableFlexibleClassical2023, sarraDeepBayesianExperimental2023, lennonEfficientlyMeasuringQuantum2019, ramseyMolecularBeamResonance1950, krantzQuantumEngineersGuide2019, hechtBeatingRamseyLimit2025, fyrillasScalableMachineLearningassisted2024}.
In experimental settings, various sources of noise influence the quantum device; therefore, it is challenging to be confident in the choice of a parametrized system model.
Alternatively, we can model the system with Blackbox models using machine learning methodologies \cite{genoisQuantumTailoredMachineLearningCharacterization2021, flynnQuantumModelLearning2022, maMachineLearningEstimation2025, youssryCharacterizationControlOpen2020, youssryExperimentalGrayboxQuantum2024, youssryModelingControlReconfigurable2020, youssryMultiaxisControlQubit2023, youssryNoiseDetectionSpectator2023, cantoneMachineLearningaidedOptimal2025, auzaQuantumControlPresence2024, mayevskyQuantumEngineeringQudits2025}. In particular, we are interested in a Graybox characterization method \cite{youssryCharacterizationControlOpen2020, youssryExperimentalGrayboxQuantum2024, youssryModelingControlReconfigurable2020, youssryMultiaxisControlQubit2023, youssryNoiseDetectionSpectator2023, cantoneMachineLearningaidedOptimal2025, auzaQuantumControlPresence2024, mayevskyQuantumEngineeringQudits2025} which is a flexible characterization method that uses a single experimental procedure for multiple realizations of a qubit.
The Graybox method models the system by combining the known mathematical procedures of the system (Whitebox) and the unknown process in the system (Blackbox). We use the experimental data to train the machine learning model to approximate the unknown process. The Whitebox and Blackbox form the Graybox predictive model. 

One of the major concerns of using deep learning is that the model can be overconfident in its predictions. Especially with a stochastic noise source, the dynamics of the system become stochastic as well. For example, a stochastic noise (colored noise) caused by an unknown Power Spectrum Density (PSD) is a noise source that is present in multiple qubit platforms \cite{sungNonGaussianNoiseSpectroscopy2019} such as superconducting, nuclear-spin, and trapped-ions qubits. Colored noise also poses a significant challenge to the construction of an accurate predictive model \cite{sunSelfconsistentNoiseCharacterization2022}, which requires specialized methods \cite{ alvarezMeasuringSpectrumColored2011, norrisQubitNoiseSpectroscopy2016, paz-silvaExtendingCombbasedSpectral2019, sungNonGaussianNoiseSpectroscopy2019,chalermpusitarakFrameBasedFilterFunctionFormalism2021} to characterize. In this work, we extend the Graybox model with Probabilistic Machine Learning \cite{p.murphyProbabilisticMachineLearning2023}, which performs inference of the model parameters using Bayesian inference. Consequently, the prediction becomes a distribution instead of a point estimate, naturally quantifying the prediction uncertainty. In particular, we implement the Blackbox part of the Graybox using \acrshort{bnn}, which can learn from a dataset without overfitting \cite{jospinHandsOnBayesianNeural2022}. Furthermore, with a probability model, it is possible to perform efficient characterization experiments with Bayesian optimal experimental design approaches \cite{fosterVariationalBayesianOptimal2019, foster2021deep}.

In this study, we use \acrfull{sgm} and our proposed \acrfull{pgm} to characterize a simulated quantum device subject to stochastic noise, i.e., colored noise.
We analyze the effect of stochastic noise through a probabilistic model of the data-generating process of a qubit. We find that the expectation value of a quantum observable (which cannot be directly observed in an experiment) becomes a distribution due to the stochastic noise. While a shifted expected value can be inferred from a finite-shot expectation value (which can be estimated from measurement data), the information about the variance of the expectation value is not accessible. Thus, the performance of the predictive model depends solely on the ability to predict the expected value of the shifted expectation value. From our experiment, we find that \acrshort{pgm} can capture the distribution of the observed data up to $\sim 1.9$ times better than the \acrshort{sgm}. Furthermore, in the control calibration task, \acrshort{pgm} can be used to find a control action that yields \acrfull{agf} closer to the optimal solution than \acrshort{sgm}. Our results enhance the Graybox characterization method by providing better uncertainty quantification through the power of Bayesian inference. We expect that our method will serve as an additional tool in constructing a reliable predictive model, enabling better control and understanding of the quantum device.

We begin by discussing the general structure of characterizing a quantum device. We will discuss the data-generating process in \cref{sec:data}, and analyze the effect of stochastic noise on the data in \cref{sec:stochastic-noise}. Next, we review the relevant details of a statistical version of the Graybox characterization method in \cref{sec:statistical-graybox} and also how to produce uncertainty associated with its prediction. Building from the foundation of \acrshort{sgm}, in \cref{sec:probabilistic-graybox}, we discuss the extension of the Graybox Characterization Method by using Probabilistic Machine Learning. Using the predictive models outlined previously, we characterize a single-qubit device subject to detuning in the X-axis and colored noise, and analyze their predictive performance in \cref{sec:characterization}. Then, we perform open-loop optimization for a quantum gate using the characterized predictive models \cref{sec:control}, and discuss their performance. We finally conclude our work in \cref{sec:conclusion}.

\section{Modeling} \label{sec:methods}

Characterization of the device aims to construct a predictive model that predicts the outcome of the actual device given control parameters, denoted by $\control$. Real physical systems are governed by laws of physics with system parameters that are only partially known by experimentalists. To address the partial knowledge of the system, one typically starts by modelling the system with a numerical model that is parametrized by model parameters. The model parameters do not necessarily have a one-to-one correspondence to the system parameters. The characterization is then performed to identify the model parameters that parametrize the predictive model using experimental data. The characterized model can then be applied to tasks of interest, such as optimal control tasks. Here, we study the case of a single-qubit predictive model, noting that it is straightforward to extend the model to larger systems, albeit at the cost of increased classical computation. The ability to predict the behavior of the device is necessary, as in realistic settings, we do not have access to the full knowledge of the system parameters.

In the following sections, we first model the data-generating process, then analyze the effect of stochastic noise on the data. We then review the mathematical construction of \acrshort{sgm}. Finally, we discuss our \acrshort{pgm} using \acrshort{bnn} built upon the foundation of \acrshort{sgm}.

\subsection{Data} \label{sec:data}

Let us start by detailing the data-generating process that models the quantum system. Consider a simple quantum system of a single qubit case with control parameters $\control$.
We note that the concept can be similarly generalized to the multiple-qubit case. The form of the control action, including the form of the function and the number of control parameters, is arbitrary.  
In our case, for the sake of demonstration, we embed the control parameters into a control signal, which is a function of time in the form $s(\control, t)$.
The total Hamiltonian governing the system is $\hat{H}_{\rm total}(s(\control, t), t)$, generating the corresponding propagator $\hat{U}_{\rm total} (\control, t)$. For simplicity, we omit the argument and define the unitary operator at the time of measurement $T$ by $\hat{U}_{\rm total} (\control) = \hat{U}_{\rm total} (\control, T)$. The expectation value of an observable $\hat{O}$ with an initial state $\rho_{0}$ is
\begin{equation} \label{eq:expectation-value}
    \langle \hat{O} \rangle_{\rho_{0}}^{\control} = \mathrm{Tr} \left[ \hat{O} \hat{U}_{\rm total}(\control) \rho_{0} \hat{U}_{\rm total}^{\dagger}(\control) \right].
\end{equation}
We note that the expectation value in \cref{eq:expectation-value} is an exact value in the case that there is no stochastic noise. However, we cannot observe  \cref{eq:expectation-value} directly in the experiment.

In the experimental setting, and focusing on the case of Pauli observables, we can obtain \cref{eq:expectation-value} only via averaging ensemble measurement, which gives binary values $b = \{ 0, 1\}$, corresponding to the eigenvalues $ e = \{ +1, -1\}$ of the quantum observable $\hat{O}$.
We have to execute $n$ identical experiments to form an ensemble average of eigenvalues $\{e_i\}_{n}$ as an estimator of the expectation value \cref{eq:expectation-value},
\begin{equation} \label{eq:finite-shot-expectation-value}
    \mathbb{E} [\hat{O}]_{\rho_{0}}^{\control} = \frac{1}{n} \sum^{n}_{i} e_{i}.
\end{equation}
We refer to the finite-shot estimation of the expectation value from $n$ bit data as the $n$-shot expectation value. The variance of this estimator is
\begin{equation}
    \mathrm{Var} (\mathbb{E} [\hat{O}]_{\rho_{0}}^{\control}) = \frac{1}{n} (1 - (\langle \hat{O} \rangle_{\rho_{0}}^{\control})^2 ),
\end{equation}
where estimation becomes exact $\mathbb{E} [\hat{O}]_{\rho_{0}}^{\control} \rightarrow  \langle \hat{O} \rangle_{\rho_{0}}^{\control}$ as $n \rightarrow \infty$. To distinguish the expectation value in \cref{eq:expectation-value} and \cref{eq:finite-shot-expectation-value}, we refer to the former as an intermediate expectation value.

\subsection{Effect of Stochastic Noise} \label{sec:stochastic-noise}

With the presence of stochastic noise, the intermediate expectation value in \cref{eq:expectation-value} is not necessarily exact and becomes a random variable. 
The intermediate expectation value represents the true performance of the control. Thus, the true performance becomes stochastic.
To illustrate the effect, we study the numerical simulation of a superconducting qubit subjected to colored noise.
We briefly explain the necessary details of the numerical study in this section. 
The control signal is given by
\begin{equation} \label{eq:signal}
    s(\mathbf{\Theta},t) = \text{Re} \left\{ h(\mathbf{\Theta}, t) e^{i(2 \pi \omega_d t + \phi)} \right\},
\end{equation}
where $h(\mathbf{\Theta}, t)$ is a control envelope, $\omega_d$ is a driving frequency, and $\phi$ is a controllable phase. Here, we consider a single control parameter $\control = \theta$ controlling the area under a Gaussian envelope with a fixed total duration of $T = 320 ~\mathrm{dt}$, a maximum possible amplitude of $A_m = 0.5$, a scale with a standard deviation $\sigma = \frac{\sqrt{2\pi}}{A_m ~ 2\pi ~ \Omega ~ \text{dt}}$ and an amplitude $A= \frac{\theta}{2\pi ~ \Omega ~ \text{dt}}$, resulting in the control envelope $h(\theta, t) = \frac{A}{\sqrt{2\pi}\sigma} \exp\left(-\frac{(t- T/2)^2}{2\sigma^2}\right)$ defined in the time step unit $\mathrm{dt} = \frac{2}{9} ~\si{ns}$.
\footnote{
    The time unit $\mathrm{dt}$ is a device sampling resolution time. This is a convention used in \texttt{qiskit} and IBM Quantum's systems \cite{qiskit2024}.
}

\begin{figure}[tb]
    \centering
    \includegraphics[width=1.\linewidth]{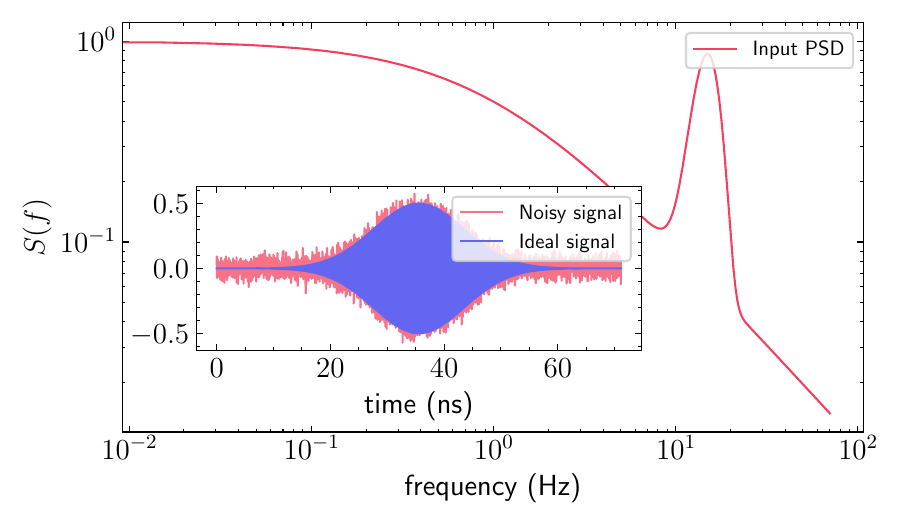}
    \caption{The Power Spectrum Density (PSD) used to generate colored noise with $\delta = 0.01$. The inset plots the ideal signal $s(2\pi, t)$ in blue and the noisy signal in red. The noisy signal produced from the sum of the ideal signal and the noise in the time-domain sample from the PSD. }
    \label{fig:psd-profile}
\end{figure}

A Power Spectrum Density (PSD) $S(f)$ is a function of frequency $f$ with units of Hz and given as
\begin{equation} \label{eq:noise-profile}
    S(f) = \frac{1}{f + 1} + 0.8 \cdot \exp\left(-\frac{(f - 15)^2}{10}\right),
\end{equation}
similar to the study in \cite{youssryCharacterizationControlOpen2020}.
To simulate noisy evolution, we use the Trotterization method and sample the noise $n(t)$ from a given PSD at each time step following the method in \cite{olivierUQpyGeneralPurpose2020}. We plot the PSD of \cref{eq:noise-profile} in \cref{fig:psd-profile}. 
The total signal sent to the system is the sum of a noiseless signal and a noise sample $s'(\control, t) = s(\control, t) + \delta n(t)$ with noise strength $\delta$. We plot a sample of a noisy signal in the inset of \cref{fig:psd-profile} with $\delta = 0.01$.

In the experiment, each noisy trajectory matches a bit string $b_{i}$. Thus, the intermediate expectation value in \cref{eq:expectation-value} becomes a random variable.
From the probability perspective, each trajectory samples the intermediate expectation value $ \langle \hat{O} \rangle_{\rho_{0}, i}^{\control}$. We do not assume any distribution form, since this is a derived random variable from noise sampled from the PSD. Next, we sample a single eigenvalue from Bernoulli distribution $e_i \sim \mathrm{Bern}((1 + \langle \hat{O} \rangle_{\rho_{0}, i}^{\control}) / 2 )$, which is equivalent to measuring the quantum circuit. Repeating this trajectory $n$ times lets us estimate the finite-shot expectation value using the estimator in \cref{eq:finite-shot-expectation-value}.


We simulate the stochastic process of controlling a quantum device using the following physical model.
Consider a system of a single qubit with frequency $\omega_q$ that is driven by a time-dependent noisy control signal $s'(\control, t)$ at strength $\Omega$. We set the driving frequency of the control signal $\omega_d = \omega_q$ in our study.
The Hamiltonian of the qubit in the rotating frame with respect to the qubit frequency is defined as
\begin{equation}
    \hat{H}_{\text{rot}} = 2 \pi \Omega \cdot s'(\mathbf{\Theta}, t) \cdot (\cos{ (2 \pi\omega_q t)} \sigma_x - \sin{ (2 \pi \omega_q t)} \sigma_y).
\end{equation}
In a realistic setting, multiple sources of noise impact the system's dynamics. However, for the simplicity and interpretability of our numerical study, we consider the case where the $\hat{H}_{\text{rot}}$ is perturbed by $\hat{H}_\mathrm{noise} = \Delta \sigma_x$ in addition to the stochastic noise given by the PSD in \cref{eq:noise-profile}, leading to the following total Hamiltonian,
\begin{equation}
    \hat{H}_{\mathrm{total}} = \hat{H}_{\text{rot}} + \hat{H}_{\text{noise}}.
\end{equation}
The noise causes the system to over-rotate, which can be corrected with the under-rotate control. The control envelope is a Gaussian envelope as defined in \cref{sec:stochastic-noise} with $\control = \theta \in [0, 2 \pi]$.
This particular choice of noise model and control action allows us to analyze the effect of noise on the system.

\begin{figure}[tb]
    \centering
    \includegraphics[width=1.\linewidth]{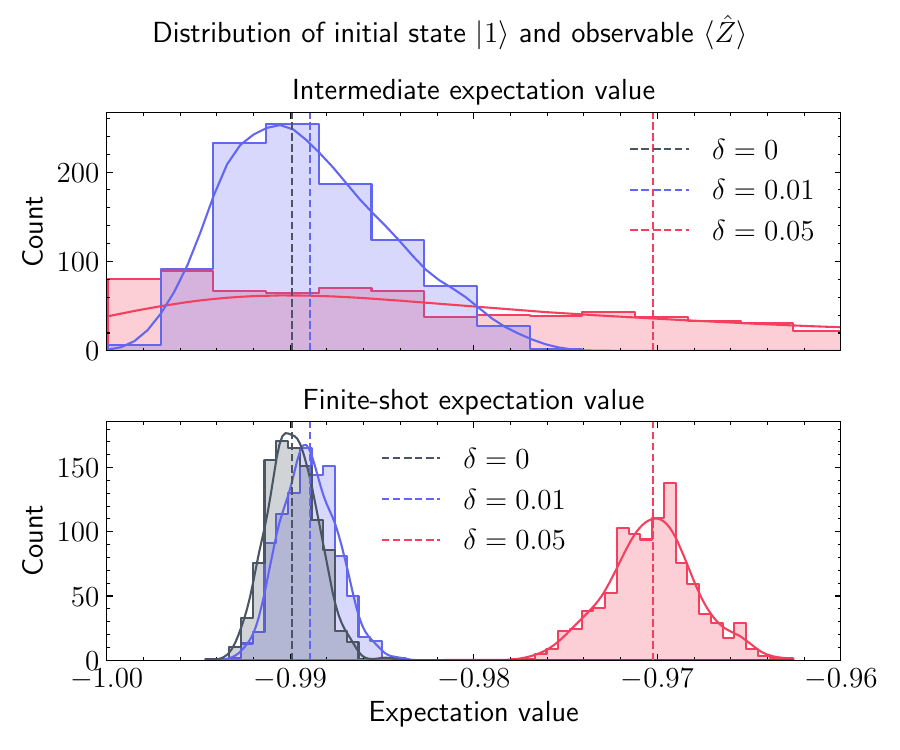}
    \caption{
        The upper plot illustrates the intermediate expectation value, which is a random variable due to the noisy signal. We consider $\delta = \{0.01, 0.05\}$, represented in blue and red colors, respectively. The gray color represented the deterministic trajectory. The vertical lines are the expected value of the distribution of the intermediate expectation value. Without noise, there is no sample for the deterministic trajectory in the upper plot. In the lower plot, we plot the samples with a size of $1,000$ of finite-shot expectation value with $n = 10,000$. We can observe that the expected value of the intermediate expectation value is approximately equal to the expected value of the finite-shot expectation value, which confirms our analytical analysis.}
    \label{fig:shift}
\end{figure}

Simulating the actual process is computationally very intensive. Thus, we approximate the process by using a resampling technique. Instead of calculating each trajectory for each shot, we approximate the distribution of intermediate expectation values with their samples, then sample with replacement from the ensemble to approximate the measurement process. To highlight the effect of stochastic noise, we simulate two noise strengths $\delta = \{ 0.01, 0.05 \}$. In addition, we also include detuning in the X-axis $\Delta \hat{\sigma}_{X}$ to the total Hamiltonian with $\Delta = 0.001$. Consider the initial state $\rho_{0} = |1\rangle \langle 1|$, observable $\hat{Z}$, and control parameter $\theta = 2 \pi$, we plot the histograms of samples of the intermediate expectation value in the upper plot of \cref{fig:shift}. We then resample from the ensemble of intermediate expectation values to produce $1,000$ finite-shot expectation values with $n = 10,000$. In the lower plot of \cref{fig:shift}, we plot the histograms of the finite-shot expectation value for both of the noise strengths and the deterministic trajectory.

To analyze the effect of the stochastic noise on the estimator in \cref{eq:finite-shot-expectation-value}, we consider the expected value and variance of the estimator in the case that the intermediate expectation value is a random variable. In the following analysis, we fix the control parameters $\control$, the initial state $\rho_0$, and the observable $\hat{O}$. Let us assume that an expected value and a variance of intermediate expectation value are $ \mathbb{E} [ \langle \hat{O} \rangle_{\rho_{0}, i}^{\control} ] = \mu_{0}$ and $ \mathrm{Var}(\langle \hat{O} \rangle_{\rho_{0}, i}^{\control}) = \sigma_{0}^2$, respectively. We refer to the expected value and variance of the intermediate expectation value as the hidden expected value and hidden variance, respectively. Since the eigenvalue is a random variable that depends on the intermediate expectation value, which is also a random variable, we must use the law of total expectation and the law of total variance to calculate the expected value and variance of the estimator of the finite-shot expectation value. We refer to the Appendix~\ref{app:estimator} for a detailed derivation.

From the law of total expectation, the expected value of the finite-shot expectation value estimator with stochastic noise is
\begin{equation} \label{eq:stochastic-expectation}
    \mathbb{E} \left[ \mathbb{E} [\hat{O}]_{\rho_{0}}^{\control} \right] = \mu_{0},
\end{equation}
which is equal to the hidden expected value. In \cref{fig:shift}, we plot the hidden expected value as vertical dashed lines. From the deterministic case in gray, the intermediate expectation value is constant, which is the expected value of the finite-shot expectation value as expected. In the case of weak stochastic noise $\delta = 0.01$ in blue, the hidden expected value shifted from the deterministic case, and became the expected value of the distribution of the finite-shot expectation value as predicted in \cref{eq:stochastic-expectation}.
In the strong stochastic noise $\delta = 0.05$ case, the shape of the distribution is not trivial. The hidden expected value shifted significantly, but the expected value of the finite-shot expectation value remains unchanged.

From the law of total variance, the variance of the finite-shot expectation value estimator with stochastic noise is
\begin{equation} \label{eq:stochastic-variance}
    \mathrm{Var} ( \mathbb{E} [\hat{O}]_{\rho_{0}}^{\control} ) = \frac{1}{n} (1 - \mu_{0}^2).
\end{equation}
The variance of the form \cref{eq:stochastic-variance} is a function of $\mu_{0}$ and the number of shots $n$ only, independent of the hidden variance. The variance of the stochastic case has the same characteristic as the variance of the estimator in the deterministic case. We can observe from the histogram of $\delta = 0.05$ presented in \cref{fig:shift}, that even with the wide hidden variance, the resulting samples of finite-shot expectation value are well-behaved as predicted in \cref{eq:stochastic-variance}.

From the expected value and variance of the estimator, we can see that the stochastic noise shifts the value of the expectation value. However, the shifted expectation value in \cref{eq:stochastic-expectation} does not contain information about the hidden variance $\sigma_{0}^2$, i.e., independent of the hidden variance. 
Consequently, we cannot quantify the complete information of the distribution of the intermediate expectation value by measuring the finite-shot expectation value. To recover the information about the noise, such as noise from a given PSD, we have to use specialized protocols \cite{alvarezMeasuringSpectrumColored2011a, chalermpusitarakFrameBasedFilterFunctionFormalism2021}.
Fortunately, accurately estimating the expected value of a finite-shot expectation value is equivalent to accurately estimating the hidden expected value, which represents the trajectory of the system evolution on average. 

\subsection{Statistical Graybox Model (SGM)} \label{sec:statistical-graybox}

\begin{figure}[tb]
    \centering
    \includegraphics[width=1.0\linewidth]{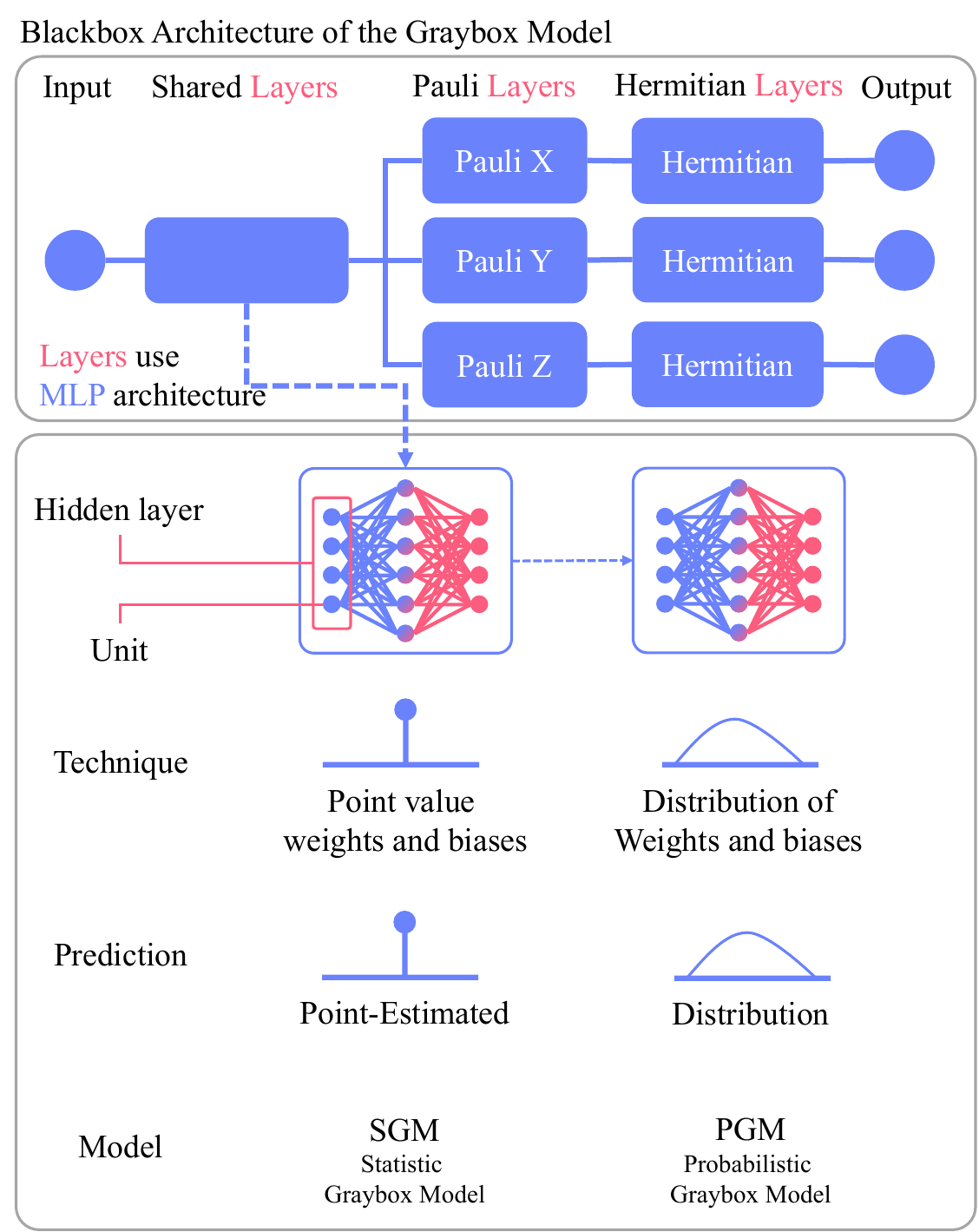}
    \caption{A high-level illustration of the Blackbox Architecture of Graybox. (1) The upper box shows a shared architecture of \acrshort{sgm}, and \acrshort{pgm}. The control is transformed by a 4th-order polynomial feature map and fed to the input layer, passing through shared layers, Pauli layers, and then converted to parameters that parametrize a Hermitian matrix for each Pauli observable. (2) The key difference of each model is the technique used in \acrshort{mlp}. (2.1) \acrshort{sgm} is a base architecture where weights and biases are point values, resulting in a deterministic point-estimated prediction. (2.2) \acrshort{pgm} implement \acrshort{bnn} which samples weights and biases from a distribution, and an ensemble of prediction forms a prediction distribution.}
    \label{fig:models-overview}
\end{figure}

The statistical Graybox characterization method models the transformation of a state as first being transformed by an ideal unitary operator $\hat{U}_0$ and then observed by a noisy observable $\hat{W}_{O}$ that deviates from the ideal observable. The expectation value predicted by the \acrshort{sgm} is given as
\begin{equation} \label{eq:exp-model}
    \mathbb{E} [\hat{O}]_{\rho_{0}}^{\control} = \mathrm{Tr} \left[ \hat{W}_{O} (\control) \hat{U}_{0} (\control)\rho_{0} \hat{U}^{\dagger}_{0} (\control) \right].
\end{equation}
The Whitebox part of \acrshort{sgm} is the ideal evolution, $\hat{U}_{0} (\control)$, obtained by solving the Schrödinger equation given an ideal parametrized Hamiltonian, $\hat{H}_{0} (\control, t)$ of the device, and the mathematical procedure leading to the expectation value.
The unknown Blackbox part, $\hat{W}_{O} (\control)$, can be modeled as a regression model that produces a Hermitian matrix of the same dimension as the system. In our case, we used \acrfull{mlp} \acrfull{dnn} to implement the Blackbox. This particular choice of architecture is not unique; however, it is simple yet flexible enough for our study.

The Deep Neural Network of Blackbox is composed of multiple \acrshort{mlp} \acrshort{dnn}s as illustrated in \cref{fig:models-overview}. The control parameters are first mapped by the function $f(x) = [x, x^2, x^3, x^4]^T$, where we substitute $f(\control / 2\pi)$ as the input to the model. The mapped input then feeds to shared layers. The output is duplicated and independently fed to Pauli layers. The number of hidden layers of shared and Pauli layers can be varied, and we chose the ReLU function as the activation function. The Hermitian Layers have a single hidden layer and are responsible for producing the output parameters $0 \leq \theta, \alpha, \beta \leq 2 \pi$, and $-1 \leq \lambda_1, \lambda_2 \leq 1 $ parameterizing $\hat{W}_{O} = \hat{U} \hat{D} \hat{U}^{\dagger}$ where,
\begin{align}
    \hat{U}(\theta, \alpha, \beta) & = \begin{pmatrix}
                                     e^{i\alpha} \cos\theta  & e^{i\beta} \sin\theta   \\
                                     -e^{-i\beta} \sin\theta & e^{-i\alpha} \cos\theta
                                 \end{pmatrix}, \\
    \hat{D}(\lambda_1, \lambda_2)  & = \begin{pmatrix}
                                     \lambda_1 & 0         \\
                                     0         & \lambda_2
                                 \end{pmatrix}.
\end{align}
A hard sigmoid activation function enforces the constraints of the output parameters, see \cite{flax2020github} for more details.

To learn the Blackbox model, an experimental dataset collected from the target system is needed.
Here, we denote the output as $\mathbf{y}$ for the observed value from the device for generality.
Typically, the dataset consists of $m$ samples. The value of $m$ is chosen to be sufficiently large so that the model can learn without overfitting.
Each sample has (input) features as parameters that parametrize the control $\control$, and the target (output) as a set of expectation values, $\mathbf{y} = \{ \mathbb{E} [\hat{O}(T)]_{\rho_{0}}^{\control} \}_{\rho_{0}, \hat{O}} $. We denote the experimental dataset as
\begin{equation}
    \mathcal{D} = \{(\control_{0}, \mathbf{y}_{0}), \ldots, (\control_{m}, \mathbf{y}_{m})\}.
\end{equation}
The complete information required to characterize the single qubit case consists of expectation values produced from combinations of $\rho_0 \in \{ |+\rangle \langle +|, |-\rangle \langle -|, |i\rangle \langle i|, |-i\rangle \langle -i|, |0\rangle \langle 0|, |1\rangle \langle 1| \}$ and $\hat{O} \in \{\hat{X}, \hat{Y}, \hat{Z}\}$. The learning algorithm is then employed to find the parameters of the model that minimize the mean square error of $K$ combinations of expectation values of the following form,
\begin{equation} \label{eq:msee-loss}
    \mathcal{L}_{\mathrm{MSE[E]}} = \frac{1}{K} \sum_{O, \rho_0} \left( \mathbb{E}[\hat{O}]^{\mathrm{model}}_{\rho_0, \control} -\mathbb{E}[\hat{O}]^{\mathrm{exp}}_{\rho_0, \control} \right)^2.
\end{equation}

Since \acrshort{sgm} is a statistical model, there is no native uncertainty corresponding to its prediction out of the box. The expectation value predicted by  \acrshort{sgm} is a point estimation. Consequently, the \acrshort{agf} is also a point estimation. However, we can quantify the uncertainty associated with the observable distribution of the \acrshort{agf}. By resampling the eigenvalue $e_{i}$ from $\mathrm{Bern} ( (1 + \langle \hat{O} \rangle_{\rho_{0}, i}^{\control} ) / 2 )$, we can form an ensemble of the finite-shot expectation value. Using the ensemble, we can calculate the samples of \acrshort{agf}. The distribution of the \acrshort{agf} naturally becomes the uncertainty. We note that the uncertainty produced by this method relies on the assumption that the predicted expectation value is the exact expectation value hidden in the experiment.

\subsection{Probabilistic Graybox Model (PGM)} \label{sec:probabilistic-graybox}

Characterization of quantum devices can be formulated within the framework of probabilistic machine learning as follows \cite{jospinHandsOnBayesianNeural2022,arbelPrimerBayesianNeural2023}. We assume that the experimental data $\mathcal{D}$ is formed by input $\mathcal{D}_{\control}$ and output $\mathcal{D}_{\mathbf{y}}$, which is generated from the likelihood $p(\mathcal{D}_{\mathbf{y}}| \mathcal{D}_{\control}, \mathbf{w})$, where $\mathbf{w}$ is a vector of model parameters characterizing the system parameters. We want to infer $\mathbf{w}$ using $\mathcal{D}$, since it contains information about the system to be characterized. From Bayes rule, the \emph{posterior distribution} is
\begin{equation} \label{eq:posterior-distribution}
    p (\mathbf{w}| \mathcal{D}) = \frac{p(\mathcal{D}_{\mathbf{y}}| \mathcal{D}_{\control}, \mathbf{w}) p(\mathbf{w})}{p(\mathcal{D})},
\end{equation}
where $p(\mathbf{w})$ is a prior distribution of $\mathbf{w}$ and $p(\mathcal{D})$ is marginal likelihood. Now, we can predict a distribution of observable  $p (\mathbf{y}^{*}| \control^{*}, \mathcal{D})$ given new control parameter $\control^{*}$ as
\begin{equation} \label{eq:posterior-predictive-distribution}
    p (\mathbf{y}^{*}| \control^{*}, \mathcal{D}) = \int p (\mathbf{y}^{*}| \control^{*}, \mathbf{w}) p (\mathbf{w}|\mathcal{D}) d \mathbf{w}.
\end{equation}
Our uncertainty over the model parameters (characterizing the system parameters) is represented by $p(\mathbf{w}|\mathcal{D})$. Each sample is then used to calculate a corresponding prediction $\mathbf{y}^{*}$, forming the \emph{posterior predictive distribution} in \cref{eq:posterior-predictive-distribution}. Since we characterize system parameters implicitly using the weights and biases of \acrshort{dnn} (Blackbox, part of \acrshort{sgm}), the statistical \acrshort{dnn} becomes \acrshort{bnn}. \acrshort{bnn} has several advantages \cite{jospinHandsOnBayesianNeural2022}. (1) It provides a natural way to quantify uncertainty. (2) It can learn from a small dataset without overfitting. (3) It is not overconfident when predicting out-of-sample data. At the same time, the Whitebox is left deterministic.

However, the closed form of the posterior distribution in \cref{eq:posterior-distribution} is generally not available. Variational Inference (VI) \cite{arbelPrimerBayesianNeural2023} is one of the methods to approximate the posterior distribution. By introducing a variational distribution $q_{\phi}(\mathbf{w})$ that is parametrized by variational parameters $\phi$, we approximate $q_{\phi}(\mathbf{w}) \approx p(\mathbf{w}|\mathcal{D})$ by finding the variational parameters that minimize the KL divergence between $q_{\phi}(\mathbf{w})$ and $p(\mathbf{w}|\mathcal{D})$. In practice, we minimize the KL divergence by maximizing an evidence lower bound (ELBO) defined as,
\begin{equation}
    \mathrm{ELBO} (\phi) = \int q_{\phi}(\mathbf{w}) \log \frac{ p(\mathbf{w}) p(\mathcal{D}| \mathbf{w}) }{q_{\phi}(\mathbf{w})} d \mathbf{w}.
\end{equation}
In our case, we use a Stochastic Variational Inference (SVI), a variant of VI, to handle a large dataset. We also minimize Trace Mean Field ELBO, which uses analytic KL divergence when possible. We refer the reader to \texttt{numpyro} documentation for more details \cite{10.5555/3322706.3322734,phanComposableEffectsFlexible2019}. The variational parameters $\phi$ are the parameters that parametrize the distribution of weights and biases of \acrshort{bnn}. In particular, we chose a normal distribution with a diagonal covariance matrix as the variational distribution. We set our prior to be a multivariate Normal distribution $\mathcal{N}(\vec{0}, 0.1 \mathbb{I})$ representing our ignorance about the true values and is numerically stable to optimize.
This transforms the inference problem into an optimization problem instead.

To perform the optimization, we must identify the stochastic process (joint distribution) that models the system's behavior as required by \texttt{numpyro}. Our stochastic model generates the observation by performing the following steps. First, given control parameters and the corresponding unitary operators $\hat{U}_{0}(\control)$ (pre-calculated using Whitebox), \acrshort{pgm} predicts the intermediate expectation value. We then use the intermediate expectation value to sample for binary measurement results. To save computational resources, for a batch of the control parameters, we sample model parameters from the variational distribution once per batch. We condition the \acrshort{pgm} with the sample of model parameters, then perform a prediction for the batch. Finally, the ELBO loss can be calculated and minimized to find the optimal variational parameters. With \acrshort{pgm}, we can predict the posterior predictive distribution, i.e., the distribution of the finite-shot expectation value given a control parameter. From the distribution, the expected value represents the prediction, while the rest of the distribution serves as a measure of uncertainty.

\section{Results} \label{sec:results}

In this section, we use \acrshort{sgm} and \acrshort{pgm} to characterize the simulated quantum device. We then use the predictive models to calibrate control parameters that maximize the \acrshort{agf} of the $\SX$ quantum gate and analyze the results.

\subsection{Characterization} \label{sec:characterization}

\begin{figure}[htb]
    \centering
    \includegraphics[width=1.0\linewidth]{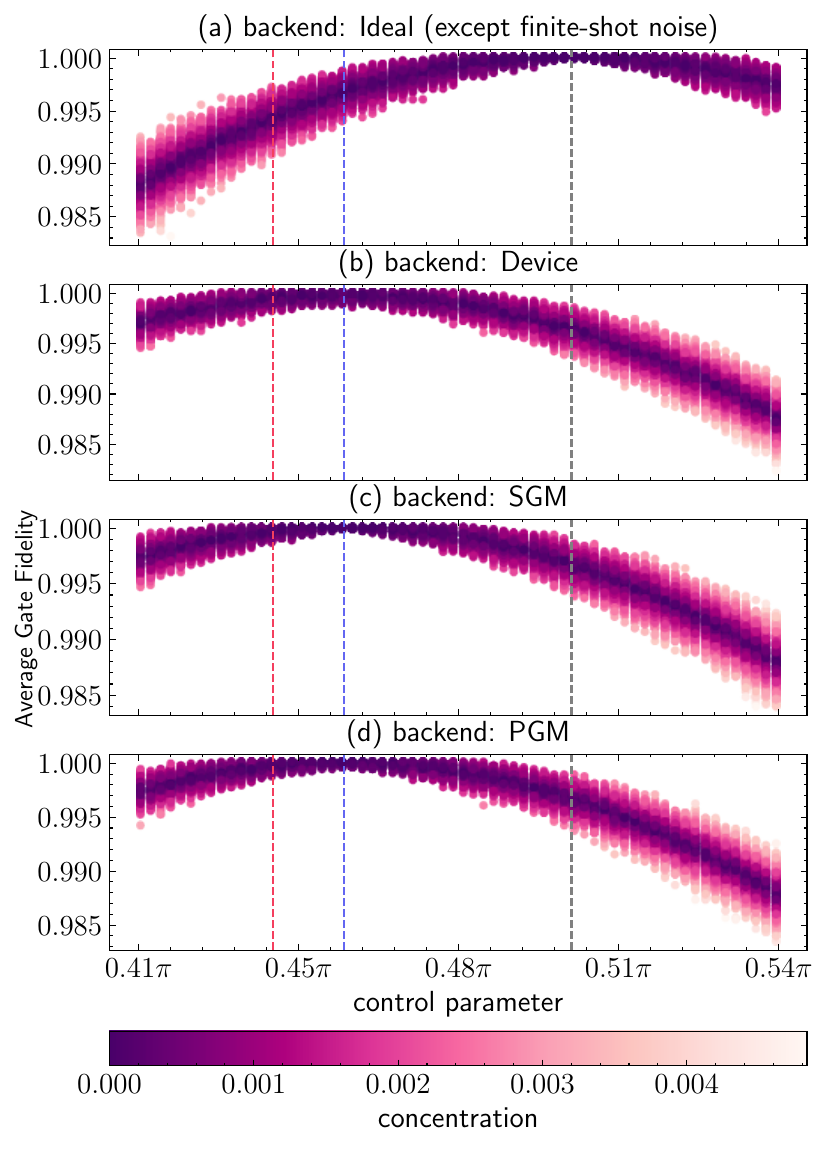}
    \caption{
    The plots visualize the distribution of the \acrshort{agf} of the $\SX$ gate predicted by different backends. Each distribution (1,000 samples) of a given control parameter is produced by assigning each sample with a concentration value that is an absolute difference from the median of the samples. We then plot each sample with various colors according to its concentration value. For each backend, we consider distributions of each value of the control parameter in a range $\theta = [ 1.3, 1.7 ]$. Figure (a) shows the AGF for the noiseless backend (simulator) except finite-shot estimation noise. Figure (b) shows the AGF for the true backend (simulator) is the device with colored, detuning noise, and finite-shot noise. Figures (c) and (d) are distributions predicted by \acrshort{sgm} and \acrshort{pgm}, respectively. The vertical dashed lines represent optimized control parameter obtained using \acrshort{sgm} in red, \acrshort{pgm} in blue, and $\theta = \pi/2$ in gray, which is the expected optimal solution without noise.
    }
    \label{fig:control-sweep}
\end{figure}

We will characterize a single qubit quantum device (simulator) using \acrshort{sgm} and \acrshort{pgm}. The device is the same device that we considered in \cref{sec:stochastic-noise} with $\delta = 0.01$. 
Before proceeding to device characterization, let us consider the behavior of the device in the case with and without noise (except noise from finite-shot estimation). 
Since we are interested in characterization for control calibration, we plot the distribution of \acrshort{agf} of the $\SX$ gate calculated from the distribution of finite-shot expectation values  executed using the simulator in \cref{fig:control-sweep}. We visualize the distribution by plotting the samples with varying colors. Given a particular control parameter, we calculate the median of the samples. For each sample, we calculate the absolute difference from the median and use the value to select the color of the sample.
The lower the value, the darker it is (closer to the median), while the higher the value, the lighter it is (farther from the median).
The \cref{fig:control-sweep}(a) shows the distributions produced by a noiseless (except finite-shot noise) device.
We can see that the \acrshort{agf} concentrates at the value of 1 at the control parameter near $\theta = \pi /2$ as expected. The vertical gray line is the control parameter at $\theta = \pi /2 $. With the presence of noise, we can see in the \cref{fig:control-sweep}(b) that the distributions of \acrshort{agf} are shifted. In any case, this particular choice of the noise model and control action allows us to identify the global optimal solution along with its distribution.

Now, we will characterize the noisy device using the Graybox method. As previously discussed, the dataset required by the Graybox characterization method consists of samples of a pair of control parameters and their corresponding combinations of expectation value.
We select a dataset of size $m = 1,000$ samples. The control parameter is sampled uniformly from the interval $ (0, 2 \pi)$.
We use our device to calculate an ensemble of intermediate expectation values given a control parameter. With the ensemble, we perform resampling to produce the finite-shot expectation values, which are the target labels of the dataset. We set the number of shots to be $n = 1,000$. With the dataset at hand, we proceed to characterize the quantum device with \acrshort{sgm} and \acrshort{pgm}.

We randomly split the dataset of size 1,000 into training and testing datasets, with sample sizes of 900 and 100, respectively. For \acrshort{sgm}, the Shared layers consist of a single dense layer of size 5, and Pauli-layers also consist of a single dense layer of size 5. Combining with the Hermitian layers, \acrshort{sgm} has 205 trainable parameters. In the case of \acrshort{pgm}, we use \acrshort{sgm} as a base model and promote it to \acrshort{bnn}, resulting in 410 variational parameters. We use \texttt{optax} \cite{deepmind2020jax} for the optimization algorithm. In particular, we use the AdamW optimizer and a learning rate schedule with cosine decay and a warm-up strategy.
For the details of the hyperparameters, see \cref{tab:detailed-hyperparams}. For the case of \acrshort{sgm}, we train the model with a mini-batch size of 100, resulting in a lower number of epochs compared to \acrshort{pgm}, which iterates through the entire training dataset in a single step. Some hyperparameters for the optimizers are shared across all experiments; otherwise, they are listed separately.
We also list hyperparameters for the optimizer used in control calibration in \cref{tab:detailed-hyperparams}. In general, we select the hyperparameters such that the model characterization finishes in a reasonable time, i.e., there is no noticeable improvement for the further optimization step. With the characterized model, we can now use the predictive models to perform predictions.

\begin{table}[htb]
    \renewcommand{\arraystretch}{1.3}
    \centering
    \begin{tabular}{llr}
        \hline\hline
        \textbf{Category}               & \textbf{Parameter}            & \textbf{Value}         \\
        \hline
        \multirow{3}{*}{Quantum System} & Qubit Frequency $\omega_{q}$  & 5.0                    \\
                                        & Qubit Drive Strength $\Omega$ & 0.1                    \\
                                        & Shots per Sample $n$          & 1000                   \\
        \hline
        \multirow{3}{*}{\makecell[l]{Noise                                                       \\Configuration}} & Detuning \(\Delta\) & 0.001                    \\
                                        & \makecell[l]{Stochastic Noise                          \\ Strength \(\delta\)} & 0.01 \\
                                        & Trotter Steps                 & 10000                  \\
        \hline
        \multirow{3}{*}{Dataset}        & Sample Size                   & 1000                   \\
                                        & Training Size                 & 900                    \\
                                        & Testing Size                  & 100                    \\
        \hline
        \multirow{5}{*}{\makecell[l]{Optimizer                                                   \\ Shared Config}} & Algorithm & adamw                    \\
                                        & Scheduler                     & \makecell{Cosine decay \\ with warmup} \\
                                        & Initial Learning Rate         & $10^{-6}$              \\
                                        & Peak Learning Rate            & 0.01                   \\
                                        & End Learning Rate             & $10^{-6}$              \\
        \hline
                                        & \textbf{Characterization}     &                        \\
        \multirow{9}{*}{\acrshort{sgm}} & \#Epoch                       & 1,000                  \\
                                        & Optimizer warm-up step        & 800                    \\
                                        & Optimizer decay step          & 8,000                  \\
                                        & \#Parameters                  & 205 (trainable)        \\
                                        & Wall Clock (s)                & $\sim$18               \\
                                        & \textbf{Control}              &                        \\
                                        & Iterations                    & 1,000                  \\
                                        & Optimizer warm-up step        & 100                    \\
                                        & Optimizer decay step          & 1,000                  \\
                                        & Wall Clock (s)                & $\sim$175              \\
        \hline
                                        & \textbf{Characterization}     &                        \\
        \multirow{9}{*}{\acrshort{pgm}} & \#Epoch                       & 10,000                 \\
                                        & Optimizer warm-up step        & 1,000                  \\
                                        & Optimizer decay step          & 10,000                 \\
                                        & \#Parameters                  & 410 (trainable)        \\
                                        & Wall Clock (s)                & $\sim$363              \\
                                        & \textbf{Control}              &                        \\
                                        & Iterations                    & 1,500                  \\
                                        & Optimizer warm-up step        & 800                    \\
                                        & Optimizer decay step          & 8,000                  \\
                                        & Wall Clock (s)                & $\sim$570              \\
        \hline
        \hline
    \end{tabular}
    \caption{Detailed System, Dataset, and Optimizer Parameters}
    \label{tab:detailed-hyperparams}
\end{table}

\begin{figure}
    \centering
    \includegraphics[width=1.0\linewidth]{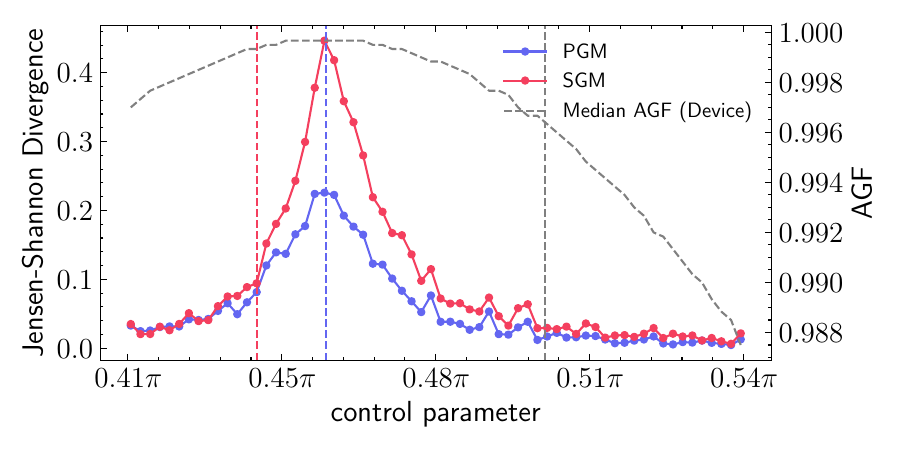}
    \caption{
    Prediction performance of \acrshort{sgm} and \acrshort{pgm} compare to the true distributions. On the left y-axis, we plot the JSD divergence (the lower the better) of distributions predicted by \acrshort{sgm} (\cref{fig:control-sweep}.(c)) in red and \acrshort{pgm} (\cref{fig:control-sweep}.(d)) in blue compare to the true distributions (\cref{fig:control-sweep}.(b)). On the right y-axis, we plot the median of \acrshort{agf} of true distribution for a reference as a gray line. We can see that the prediction performance of both models drops when the control parameters approach the optimal solution. The vertical dashed lines are (1) optimized control parameter optimized using \acrshort{sgm} in red, (2) optimized control parameter optimized using \acrshort{pgm} in blue, and $\theta = \pi/2$ in gray which is the expected optimal solution without noise.
    }
    \label{fig:jsd-agf}
\end{figure}

To demonstrate the performance of the predictive model, we also plot the distribution of \acrshort{agf} predicted by \acrshort{sgm} and \acrshort{pgm} in \cref{fig:control-sweep}(c) and \cref{fig:control-sweep}(d), respectively. However, it is difficult to visually distinguish the differences between the distributions predicted by each backend. We compare the closeness of the two distributions with the Jensen-Shannon Divergence distance \cite{nielsenJensenShannonSymmetrization2019}.  We choose the JSD over the Kullback-Leibler divergence for its numerical stability. JSD is bounded within $[0, \ln(2)]$ \cite{nielsenJensenShannonSymmetrization2019}, where lower is better (two distributions are close to each other). JSD is also symmetric in its arguments. Since the data we have are empirical samples, we use a binning strategy for the calculation. We plot the JSD for each control parameter in \cref{fig:jsd-agf} with the same interval in \cref{fig:control-sweep} using \acrshort{sgm} (red) and \acrshort{pgm} (blue). We also plot the median of \acrshort{agf} simulated by the device as a dashed gray line. Observe that both predictive models have poor performance near the control parameter value that yields the \acrshort{agf} near optimal value. This is the consequence of choosing the $\SX$ gate. Since the expectation values that maximize the \acrshort{agf} of $\SX$ gate have a value of $\{ -1, 0, 1 \}$, it is harder for a predictive model to predict a distribution with near-zero variance. However, we clearly observe that \acrshort{pgm} performs substantially better than \acrshort{sgm}, particularly at the optimal control parameter.

\subsection{Control Calibration} \label{sec:control}

To demonstrate the control inference process, we infer the control parameter of the $\SX$ gate with a number of shots $n = 1000$ shots. We note that this particular choice is arbitrary. A sequence of $\SX$ and a rotation around the Z-axis gate can form a universal single-qubit gate \cite{mckayEfficientGatesQuantum2017}.
In the statistical regime, when \acrshort{sgm} is ready, one can perform control calibration with either a model-based or model-free approach to obtain control parameters that maximize performance metrics. Especially with the \acrshort{sgm}, one can perform gradient-based optimization to accomplish the task.
For \acrshort{sgm}, we minimize the cost function $(1 - \mathrm{AGF})^{2}$ using optimizer as detailed in \cref{tab:detailed-hyperparams}. We calculate \acrshort{agf} using the intermediate expectation value predicted by \acrshort{sgm} directly instead of using finite-shot expectation value. Thus, optimization is deterministic in \acrshort{sgm} case. 
In the case of \acrshort{pgm}, we reframe the problem into inferring the control parameters that produce the desired distributions, i.e., finding the control parameters that produce the distributions of the ideal target control.
We achieve inference using \acrfull{mle} \cite{10.5555/3322706.3322734}.
The resulting distribution of the finite-shot expectation value can be used to calculate performance metrics for further analysis. We list the optimized parameters by each predictive model in \cref{tab:control-calibration-stat}. For each optimized control, we use the device, \acrshort{sgm}, and \acrshort{pgm} to predict the \acrshort{agf} and list their expected values in \cref{tab:control-calibration-stat}. We also plot vertical dashed lines corresponding to the optimized control parameters by \acrshort{sgm} $\theta_\mathrm{SGM}^{*}$ (red) and \acrshort{pgm} $\theta_\mathrm{PGM}^{*}$ (blue) in \cref{fig:control-sweep} and \cref{fig:jsd-agf}.

\begin{table}[htb]
    \renewcommand{\arraystretch}{1.3}
    \centering
    \begin{tabular}{llr}
        \hline\hline
        \textbf{Control} & \textbf{Parameter}                        & \textbf{Value}  \\
        \hline
        \multirow{7}{*}{\acrshort{sgm}}
                         & $\theta^{*}$                              & 1.384006        \\
                         & $\mathbb{E}[\mathrm{AGF}]$ by True        & 0.9993480       \\
                         & $\mathbb{E}[\mathrm{AGF}]$ by SGM         & 0.9996295       \\
                         & $\mathbb{E}[\mathrm{AGF}]$ by PGM         & 0.9996110       \\
                         & $D_{\rm JSD}(\mathrm{SGM}|\mathrm{True})$ & 0.1064          \\
                         & $D_{\rm JSD}(\mathrm{PGM}|\mathrm{True})$ & 0.0920          \\
                         & Ratio                                     & \textbf{1.1568} \\
        \hline
        \multirow{7}{*}{\acrshort{pgm}}
                         & $\theta^{*}$                              & 1.428679        \\
                         & $\mathbb{E}[\mathrm{AGF}]$ by True        & 0.9996952       \\
                         & $\mathbb{E}[\mathrm{AGF}]$ by SGM         & 0.9999978       \\
                         & $\mathbb{E}[\mathrm{AGF}]$ by PGM         & 0.9999430       \\
                         & $D_{\rm JSD}(\mathrm{SGM}|\mathrm{True})$ & 0.4404          \\
                         & $D_{\rm JSD}(\mathrm{PGM}|\mathrm{True})$ & 0.2315          \\
                         & Ratio                                     & \textbf{1.9024} \\
        \hline
        \hline
    \end{tabular}
    \caption{Statistics relevant to the control calibration experiments}
    \label{tab:control-calibration-stat}
\end{table}

\begin{figure*}[htb]
    \centering
    \subfloat[$\theta_\mathrm{SGM}^{*}$]{\label{fig:cc-sgm}
        \includegraphics[width=.480\linewidth]{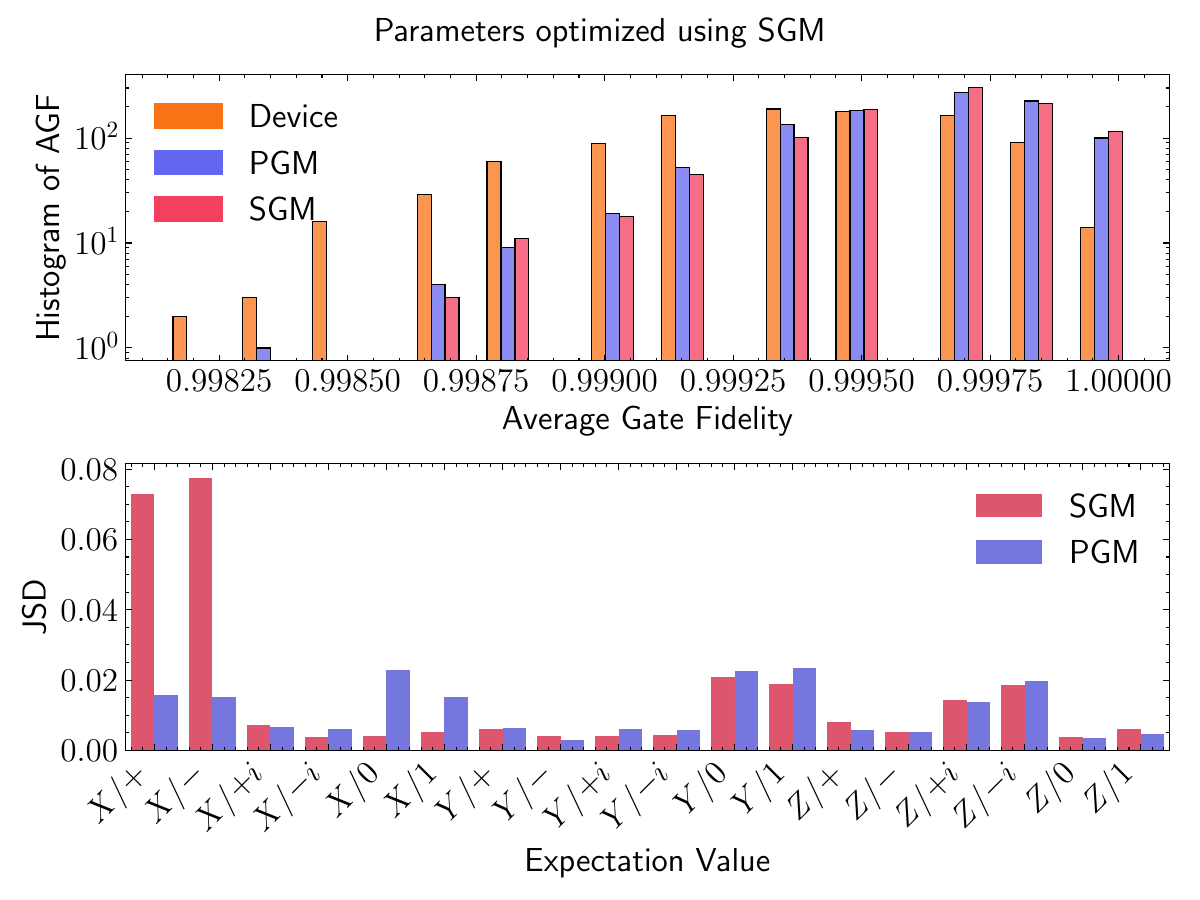}}
    \subfloat[$\theta_\mathrm{PGM}^{*}$]{
        \label{fig:cc-pgm}
        \includegraphics[width=.480\linewidth]{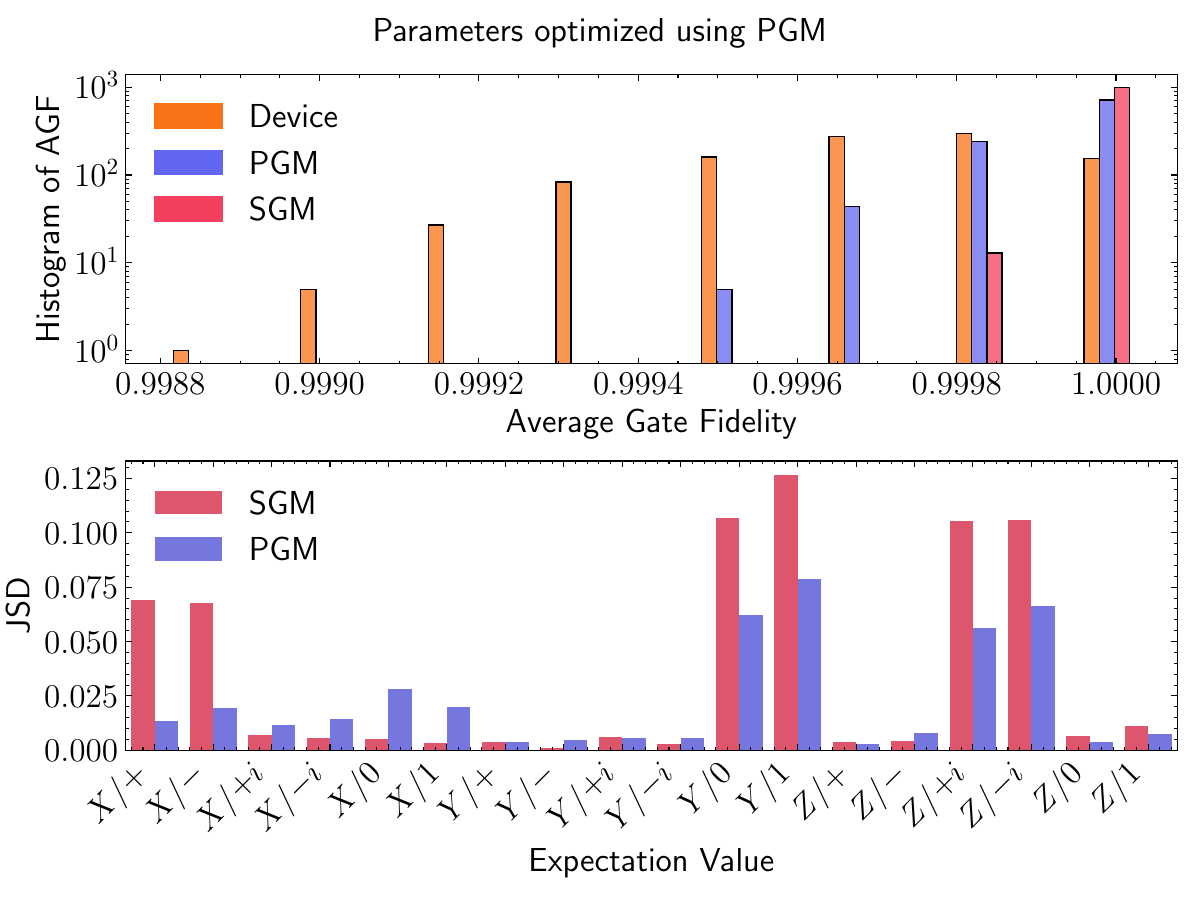}
    }
    \caption{
        On the top panels are plot of the histogram of 1,000 realizations of \acrshort{agf} of $\SX$ gate calculation using ground truth (target) simulator and prediction from \acrshort{sgm} and \acrshort{pgm}. On lower panels, we plot the bar compare the JSD between the distribution of each expectation value predict by both predictive model compare to the distribution of the device.}
    \label{fig:cc-hist}
\end{figure*}

However, reporting expected value and variance of \acrshort{agf} is not informative since the distribution of \acrshort{agf} is not necessarily Gaussian. To demonstrate the distributions of \acrshort{agf}, we plot the histogram of samples of \acrshort{agf} in the upper panels of \cref{fig:cc-hist}. For each optimized parameter, we compare the histograms of samples predicted by the device (orange), \acrshort{sgm} (red), and \acrshort{pgm} (blue). Quantitatively, \acrshort{pgm} predicts a distribution closer to the device than \acrshort{sgm}. From \cref{tab:control-calibration-stat}, the ratio of JSD predicted by \acrshort{sgm} over \acrshort{pgm} is 1.3055 (approximately $\sim 1.3$) in the case of the optimized parameter predicted by \acrshort{sgm} and 1.9024 (approximately $\sim 1.9$) in the case of the optimized parameter predicted by \acrshort{pgm}.
From \cref{fig:control-sweep}, $\theta_\mathrm{PGM}^{*}$ is close to the optimal solution already. The poor prediction performance at $\langle \hat{X} \rangle_{|+\rangle}, \langle \hat{X} \rangle_{|-\rangle}, \langle \hat{Y} \rangle_{|0\rangle}, \langle \hat{Y} \rangle_{|1\rangle}, \langle \hat{Z} \rangle_{|r\rangle}, \langle \hat{Z} \rangle_{|l\rangle}$ is because they require samples that are distributed close to a sharp Delta distribution at eigenvalue of $e_{i} = \{-1, 1\}$, which have a very narrow variance, thus a slight deviation of expected value results in a large value of JSD.
Thus, the choice of $\SX$ reveals the importance of a predictive model capable of handling such a case.
While \acrshort{sgm} is capable of detecting a shift of the data and produce the distribution in a similar shape as \acrshort{pgm} as shown in \cref{fig:control-sweep}, $\theta_\mathrm{SGM}^{*}$ is farther from the global optimal solution. We can see from \cref{tab:control-calibration-stat} that the solution yields good performance. The suboptimal parameters might result from a choice of optimizer hyperparameters.

The numerical results suggest that \acrshort{pgm} performs better than \acrshort{sgm}. However, we would like to note again that our simulation uses approximation at multiple stages. First, we approximated the propagator using Trotterization and approximated the distribution of intermediate expectation value with a smaller sample size (resample technique). These approximation steps might introduce deviation from the true distribution. Nonetheless, we can see that given the same dataset and our setting, \acrshort{pgm} is more reliable than \acrshort{sgm} in device characterization and control calibration tasks, especially when expectation values yielded from the target quantum gate have concentrated distributions such as $\SX$ chosen in this study.

\section{Conclusion} \label{sec:conclusion}

We discussed how to augment the Graybox characterization method with uncertainty quantification. First, we introduced the original version of the Graybox model and how to obtain its prediction uncertainty. Second, we used probabilistic machine learning, equipping the method with a natural ability to quantify uncertainty. Our results show that \acrshort{pgm} can capture the distribution of the observed data better than \acrshort{sgm} up to $\sim1.9$ times.
We reframed a problem of control calibration as a maximum likelihood estimation problem and utilized the probabilistic Graybox model to calibrate the $\SX$ gate, comparing the result with its statistical counterparts. The control parameter predicted by \acrshort{pgm} is closer to the global optimal solution than the \acrshort{sgm}'s prediction. 
Our analysis reveals that the performance of the predictive model for the quantum device depends only on the accuracy of the expected value of the expectation value prediction.
We envision our proposed \acrshort{pgm} will be a valuable tool for understanding the behavior of a quantum device and for calibrating quantum operations. 

Our analysis of the stochastic noise effect on data, characterization, and control calibration of the $\SX$ gate suggests that developing a predictive model capable of handling an extremum case of distribution expectation value would be a valuable future work.
The extension of PGM is similar to the extension of SGM, since they share the same mathematical formulation. For example, in the case of a two-or-more-qubit system, we must measure a complete set of combinations of expectation values, i.e., a tomographically complete set for process characterization. In the \cite{mayevskyQuantumEngineeringQudits2025}, they studied the extension of the method to the qudits system.

\section*{Data Availability}
The code used to produce the results in this study is available at https://github.com/PorametPat/bnn-graybox.git.

\begin{appendices}

    \section{Expected value and variance of the estimator} \label{app:estimator}

    We show a detailed derivation of the expected value of the estimator of the finite-shot expectation value \cref{eq:stochastic-expectation}. The expected value of the estimator in \cref{eq:finite-shot-expectation-value} can be written as,
    \begin{equation}
        \mathbb{E} \left[ \mathbb{E} [\hat{O}]_{\rho_{0}}^{\control} \right] = \frac{1}{n} \sum_{i}^{n} \mathbb{E} [e_{i}].
    \end{equation}
    Since $e_{i}$ is a random variable that depends on another random variable, we must use the law of the total expectation as follows,
    \begin{equation}
        \mathbb{E}[e_{i}] = \mathbb{E} \left[ \mathbb{E} \left[ e_{i} | \langle \hat{O} \rangle_{\rho_{0}, i}^{\control} \right] \right].
    \end{equation}
    Consider the inner expected value, which is the expected value of $e_{i}$ condition on the intermediate expectation value $\langle \hat{O} \rangle_{\rho_{0}, i}^{\control}$. The eigenvalue $e_{i}$ is sample from $e_i \sim \mathrm{Bern}((1 + \langle \hat{O} \rangle_{\rho_{0}, i}^{\control}) / 2 )$.
    The conditional expected value is simplified to
    \begin{equation}
        \mathbb{E} \left[ e_{i} | \langle \hat{O} \rangle_{\rho_{0}, i}^{\control} \right] = \langle \hat{O} \rangle_{\rho_{0}, i}^{\control}.
    \end{equation}
    Substituting the conditional expectation back to the total expectation, we can simplify the total expectation by using the assumption that $\mathbb{E} [ \langle \hat{O} \rangle_{\rho_{0}, i}^{\control} ] = \mu_{0}$ which yield the total expectation of the form
    \begin{equation}
        \mathbb{E}[e_{i}] = \mathbb{E} [ \langle \hat{O} \rangle_{\rho_{0}, i}^{\control} ] = \mu_{0}.
    \end{equation}

    Next, we consider the variance of the estimator,
    \begin{equation}
        \mathrm{Var} \left( \mathbb{E} [\hat{O}]_{\rho_{0}}^{\control} \right) = \frac{1}{n^2} \sum_{i}^{n} \mathrm{Var} ( e_{i} ).
    \end{equation}
    In a similar manner, we must calculate the variance of the estimator using the law of total variance as follows,
    \begin{equation} \label{eq:total-variance}
        \mathrm{Var} ( e_{i} ) = \mathbb{E}[ \mathrm{Var} ( e_{i} | \langle \hat{O} \rangle_{\rho_{0}, i}^{\control} ) ] + \mathrm{Var} \left( \mathbb{E} \left[ e_{i} | \langle \hat{O} \rangle_{\rho_{0}, i}^{\control} \right]  \right).
    \end{equation}
    Consider the conditional variance in the first term. We expand and rewrite it in the following,
    \begin{align}
        \mathrm{Var} ( e_{i} | \langle \hat{O} \rangle_{\rho_{0}, i}^{\control} ) & = \mathbb{E} \left[ \left( e_{i} - \mathbb{E} \left[e_{i} | \langle \hat{O} \rangle_{\rho_{0}, i}^{\control} \right] \right)^2 | \langle \hat{O} \rangle_{\rho_{0}, i}^{\control} \right] \\
                                                                                  & = \mathbb{E} \left[e_{i}^2 | \langle \hat{O} \rangle_{\rho_{0}, i}^{\control} \right] -\left( \langle \hat{O} \rangle_{\rho_{0}, i}^{\control} \right)^2 \label{eq:last-1}
    \end{align}
    Since, we know the form of the distribution of the eigenvalue $e_{i}$, we can simplify the first term in \cref{eq:last-1} to $\mathbb{E} \left[e_{i}^2 | \langle \hat{O} \rangle_{\rho_{0}, i}^{\control} \right] = 1$. Thus, the conditional variance becomes
    \begin{equation}
        \mathrm{Var} ( e_{i} | \langle \hat{O} \rangle_{\rho_{0}, i}^{\control} ) = 1 - \left( \langle \hat{O} \rangle_{\rho_{0}, i}^{\control} \right)^2,
    \end{equation}
    which is the variance of a quantum observable $\hat{O}$ given that $\hat{O}$ is Hermitian. For the second term of \cref{eq:total-variance}, we can use what was derived previously, thus the total variance becomes,
    \begin{equation}\mathrm{Var} ( e_{i} ) = 1 - \left( \langle \hat{O} \rangle_{\rho_{0}, i}^{\control} \right)^2 + \mathrm{Var} \left( \langle \hat{O} \rangle_{\rho_{0}, i}^{\control}\right).
    \end{equation}
    Using the definition of variance of a random variable, we can rewrite the total variance as follows,
    \begin{equation}
        \mathrm{Var} ( e_{i} ) = 1 - \left(  \mathbb{E} \left[ \langle \hat{O} \rangle_{\rho_{0}, i}^{\control} \right]  \right)^2 = 1 - \mu_{0}^{2}.
    \end{equation}
    Finally, the variance of the estimator becomes,
    \begin{equation}
        \mathrm{Var} \left( \mathbb{E} [\hat{O}]_{\rho_{0}}^{\control} \right) = \frac{1}{n^2} n (1 - \mu_{0}^2) = \frac{1}{n}(1 - \mu_{0}^2).
    \end{equation}
    Showing that the variance of the estimator is independent of the variance of the intermediate expectation value as discussed in the \cref{sec:stochastic-noise}.

\end{appendices}

\section*{Acknowledgment}
We would like to thank Areeya Chantasri, Theerapat Tansuwannont and Kaiah Steven for fruitful discussion. In this work, we acknowledge the use of Raycast AI to help format the tables. 

\bibliographystyle{IEEEtran}
\bibliography{references}

\EOD

\end{document}